\documentclass[unnumsec,webpdf,contemporary,large]{oup-authoring-template}
\graphicspath{{Fig/}}
\usepackage{amsmath}
\usepackage{graphicx}
\usepackage{hyperref}
\usepackage{lineno}
\usepackage{float}
\usepackage{microtype}
\usepackage{rotating} 
\theoremstyle{thmstyleone}

\theoremstyle{thmstylethree}

\hypersetup{
    colorlinks=true,
    linkcolor=blue,
    citecolor=blue,
    urlcolor=blue
}

\journaltitle{Briefings in Bioinformatics} 
\DOI{in press} 
\copyrightyear{2025} 
\pubyear{2025} 
\access{Advance Access Publication Date: TBD} 
\appnotes{Paper} 
\firstpage{1}

\title[Predicting AMR in \textit{Campylobacter}]{Predicting Antimicrobial Resistance (AMR) in \textit{Campylobacter}, a Foodborne Pathogen, and Cost Burden Analysis Using Machine Learning}
\author[1,$\ast$]{Shubham Mishra\ORCID{https://orcid.org/0009-0004-2648-8814}}
\author[1]{The Anh Han\ORCID{https://orcid.org/0000-0002-3095-7714}}
\author[2,3]{Bruno Silvester Lopes\ORCID{https://orcid.org/0000-0002-1476-2108}}
\author[1]{Shatha Ghareeb\ORCID{https://orcid.org/0009-0006-4177-3562}}
\author[1]{Zia Ush Shamszaman\ORCID{https://orcid.org/0000-0003-1954-1950}}
\authormark{Shubham Mishra et al.}

\address[1]{\orgdiv{School of Computing, Engineering and Digital Technologies}, \orgname{Teesside University}, \orgaddress{\state{Middlesbrough}, \country{United Kingdom}, \postcode{TS1 3BX}}}
\address[2]{\orgdiv{School of Health and Life Sciences}, \orgname{Teesside University}, \orgaddress{\state{Middlesbrough}, \country{United Kingdom}, \postcode{TS1 3BX}}}
\address[3]{\orgdiv{National Horizons Centre}, \orgname{Teesside University}, \orgaddress{\state{Darlington}, \country{United Kingdom}, \postcode{DL1 1HG}}}

\corresp[$\ast$]{Corresponding author. \href{email:z.shamszaman@tees.ac.uk}{z.shamszaman@tees.ac.uk}}

\received{TBD}{0}{TBD}
\revised{TBD}{0}{TBD}
\accepted{TBD}{0}{TBD}

\abstract{
Antimicrobial resistance (AMR) poses a significant public health and economic challenge, increasing treatment costs and reducing antibiotic effectiveness. This study employs machine learning to analyze genomic and epidemiological data from the public databases for molecular typing and microbial genome diversity (PubMLST), incorporating data from UK government-supported Antimicrobial Resistance (AMR) surveillance by the Food Standards Agency and Food Standards Scotland, identifying AMR patterns in \textit{Campylobacter jejuni} and \textit{Campylobacter coli} isolates from the UK, spanning 2001--2017. The research integrates whole-genome sequencing (WGS) data, epidemiological metadata, and economic projections to identify key resistance determinants and forecast future resistance trends and healthcare costs. The study investigates \textit{gyrA} mutation for fluoroquinolone resistance and the \textit{tet(O)} gene for tetracycline resistance, training a Random Forest (RF) model, validated with bootstrap resampling (1,000 samples, 95\% confidence intervals), achieving 74\% accuracy in predicting AMR phenotypes. Time-series forecasting models, namely, Seasonal Autoregressive Integrated Moving Average (SARIMA), Susceptible-Infected-Recovered (SIR), and Prophet, predict a rise in campylobacteriosis cases, potentially exceeding 130 cases per 100,000 by 2050, with an economic burden surpassing £1.9 billion annually if unchecked. An enhanced Random Forest-based system, analyzing 6,683 isolates, refines predictions by incorporating temporal patterns, uncertainty estimation, and resistance trend modeling up to 2050, indicating sustained high $\beta$-lactam resistance (~100\%), rising fluoroquinolone resistance, and fluctuating tetracycline resistance.
}

\keywords{antimicrobial resistance, machine learning, \textit{Campylobacter}, genomics, random forest, time-series forecasting, economic burden}

\begin{document}
\maketitle
\section{Introduction}
Antimicrobial resistance (AMR) represents one of the most significant challenges to global public health in the 21st century. The World Health Organization has identified AMR as a priority health threat, with estimates suggesting that resistant infections could cause 10 million deaths annually by 2050 \cite{who2019}. A comprehensive global analysis revealed that \textit{Campylobacter} species were responsible for approximately 75,000 deaths attributable to AMR in 2019, making them one of the leading foodborne bacterial pathogens contributing to AMR mortality \cite{murray2023}. Among foodborne pathogens, \textit{Campylobacter} ranks second only to \textit{Escherichia coli} in terms of AMR-associated deaths globally \cite{murray2023}. Among the pathogens of concern, \textit{Campylobacter} species, particularly \textit{C. jejuni} and \textit{C. coli}, present a significant burden due to their prevalence in foodborne illness and increasing resistance to frontline antibiotics \cite{efsa2024}.

\textit{Campylobacter} is the leading cause of bacterial gastroenteritis in the United Kingdom (UK) and many developed countries, with approximately 630,000 cases reported annually in the UK alone \cite{fsa2017}. The economic impact of these infections is substantial, encompassing direct healthcare costs, productivity losses, and long-term sequelae such as Guillain-Barré syndrome and reactive arthritis \cite{mangen2015}. As resistance to fluoroquinolones, tetracyclines, and $\beta$-lactams continues to rise, effective treatment options are diminishing, exacerbating the clinical and economic burden \cite{luangtongkum2009}.

Traditional approaches to AMR surveillance rely heavily on phenotypic susceptibility testing, which is time-consuming and resource-intensive \cite{efron2016}. Recent advances in whole-genome sequencing (WGS) and machine learning present opportunities to revolutionize AMR prediction and surveillance by identifying genomic markers associated with resistance phenotypes \cite{nguyen2019}. These computational approaches can potentially accelerate detection, improve accuracy, and enable large-scale monitoring of resistance trends \cite{moradigaravand2018}.

This study develops and validates a machine learning framework for predicting AMR in \textit{Campylobacter} species using genomic and epidemiological data, and to forecast future resistance trends and associated economic impacts. By integrating molecular, phenotypic, and economic analyses, we seek to provide a comprehensive understanding of the AMR landscape and inform evidence-based policies for antimicrobial stewardship.

\section{Related Works}

\subsection{Genomics and AMR}
Several studies have demonstrated the utility of whole-genome sequencing (WGS) data for AMR prediction. \cite{moradigaravand2018} developed a random forest model to predict antibiotic resistance in \textit{Escherichia coli} using pan-genome data, achieving accuracies exceeding 90\% for multiple antibiotics. Their approach highlighted the importance of accessory genome elements in conferring resistance, beyond well-characterised resistance genes.

\cite{lopes2019} tracked the emergence and evolution of multidrug-resistant \textit{Campylobacter jejuni} sequence type 5136 in the UK, demonstrating how WGS can reveal the development and spread of resistant lineages. Their longitudinal analysis revealed progressive acquisition of resistance determinants and their temporal distribution, informing surveillance strategies.

\subsection{Machine Learning in AMR}
Advanced computational methods have significantly enhanced AMR prediction capabilities. \cite{nguyen2019} applied machine learning algorithms to predict minimum inhibitory concentrations (MICs) for nontyphoidal \textit{Salmonella}, demonstrating that ensemble methods could accurately forecast resistance phenotypes based on genomic features. Their approach integrated multiple algorithms, including random forests and gradient boosting, to optimize predictive performance.

\cite{arango-argoty2018} developed DeepARG, a deep learning approach for predicting antibiotic resistance genes from metagenomic data. This method addressed limitations in homology-based approaches, improving detection of novel resistance determinants in complex microbial communities. Their work demonstrated the potential of neural networks to capture subtle genomic signatures associated with resistance.

\subsection{Forecasting Methods}
Temporal modeling of AMR patterns represents another important research direction. \cite{niewiadomska2019} conducted a systematic review of population-level mathematical modeling approaches for AMR, identifying key methodological considerations for forecasting resistance trends. Their analysis highlighted the importance of incorporating uncertainty quantification and model validation in long-term projections.

The principles for time-series forecasting that have been adapted for AMR prediction, emphasizing the importance of handling seasonality and long-term trends in resistance patterns \cite{hyndman2006}. Their methodological framework provides a foundation for the time-series approaches employed in this study.

\subsection{Economic Burden Analysis}
The economic impact of AMR has been investigated through various modeling approaches. \cite{smith2013} estimated the true cost of antimicrobial resistance, incorporating both direct healthcare expenses and broader societal impacts. Their work highlighted methodological considerations for comprehensive economic assessment of AMR.

Building on this foundation, \cite{jit2020} proposed a conceptual framework for quantifying the economic cost of antibiotic resistance and evaluating intervention impacts. Their approach integrated healthcare costs, productivity losses, and mortality effects into a unified economic model, providing valuable guidance for the cost analysis component of this study.

\cite{oneill2016} landmark report on tackling drug-resistant infections outlined the potential global economic impact of unchecked AMR, projecting cumulative costs of \$100 trillion by 2050 if current resistance trends continue. This comprehensive analysis established the macroeconomic context for more targeted studies of pathogen-specific economic burdens.

\section{Materials and Methods}

\subsection{Data Description}
This study used a dataset comprising 6,683 whole-genome sequenced \textit{Campylobacter jejuni} and \textit{C. coli} isolates, collected between 2001 and 2017 from clinical and animal sources in the UK. The dataset included genotypic information (AMR genes, key mutations, Multilocus Sequence Typing profiles), phenotypic resistance profiles, and epidemiological metadata (host, region, isolation year, and source). Genomic data were obtained from the public databases for molecular typing and microbial genome diversity (PubMLST) and were part of the high profile reports published earlier by Food Standards Agency/Food Standards Scotland which inform policies in the UK related to food security and safety \cite{icamps2017,campyAttribution2017}.

It comprises genotypic data (\textit{gyrA} mutations, \textit{tet(O)}, $\beta$-lactamase genes like \textit{blaOXA-61} and \textit{OXA-184}), phenotypic resistance profiles, and metadata (host, region, year, source). The preprocessing, performed using Python with Pandas and Scikit-learn, removed duplicates, imputed missing values, and introduced temporal features (“years since 2000”), allowing the models to capture long-term progression of AMR.

\subsection{Model Design}

\subsubsection{ML Framework for AMR Phenotype Prediction}
A random forest classifier was used to predict AMR phenotypes based on genomic and epidemiological characteristics, following the approach of \cite{moradigaravand2018}. The model was trained to classify resistance across three antibiotic classes:
\begin{itemize}
    \item Fluoroquinolones (FQ): Predicted using \textit{gyrA} mutations (T86I), which are well-established markers of quinolone resistance \cite{Payot2006}.
    \item Tetracyclines (Tet): Predicted using \textit{tet(O)} and other \textit{tet(O)}-like variants.
    \item $\beta$-Lactams (BL): Predicted using $\beta$-lactamase gene presence like \textit{blaOXA-61} and \textit{OXA-184} for BL \cite{Griggs2009}.
\end{itemize}

Feature selection was performed using recursive feature elimination (RFE) to identify key genomic and epidemiological predictors, as recommended earlier \cite{Kuhn2019}. RFE was compared with other methods such as Boruta and Mutual Information, finding that RFE provided a balanced set of features with high predictive power. The model was optimised using grid search combined with 5-fold cross-validation to reliably enhance accuracy and reducing the risk of overfitting. The final configuration was set to 100 decision trees, a maximum depth of 10, and a minimum leaf size of 5.

\begin{sidewaystable*}[htbp]
\small
\begin{tabular}{@{\extracolsep\fill}l p{4cm} p{6cm} p{6cm} p{6cm}@{\extracolsep\fill}}
\toprule
\textbf{Method} & \textbf{Description} & \textbf{Advantages} & \textbf{Disadvantages} \\
\midrule
RFE & Removes weakest features iteratively & Efficient, RF integration & May miss interactions \\
Boruta & Compares with shadow features & Robust, noise handling & Computationally intensive \\
Mutual Information & Measures target dependency & Non-linear relationships & High-dimensional intensity \\
\botrule
\end{tabular}
\caption{Comparison of feature selection methods (RFE, Boruta, and Mutual Information) with their descriptions, advantages, and disadvantages.}
\label{tab:feature-selection}
\end{sidewaystable*}

\subsubsection{Time-Series Forecasting Models for AMR Trends}
To predict future resistance prevalence and clinical burden, the study employed three forecasting models, each chosen for their strengths in handling different aspects of time-series data:
\begin{enumerate}
    \item \textbf{SARIMA (Seasonal Autoregressive Integrated Moving Average)}: This model is effective for capturing seasonal patterns and trends in time-series data. It can model complex seasonal fluctuations and autocorrelation but requires the data to be stationary and may not handle non-linear trends well \cite{Hyndman2021}.
    \item \textbf{Prophet Model}: Developed by Facebook, Prophet is designed for forecasting time series with strong seasonal effects and missing data. It is robust to outliers and provides uncertainty intervals, making it suitable for long-term trend forecasting \cite{Taylor2018}.
    \item \textbf{SIR (Susceptible-Infected-Recovered) Epidemiological Model}: This model integrates AMR resistance rates to estimate future infection burden and healthcare costs. It incorporates disease dynamics and can simulate the impact of interventions, though it relies on accurate parameter estimation \cite{niewiadomska2019}.
\end{enumerate}

These models were trained on historical data from 2001 to 2017 and used to forecast trends up to 2050.

\subsubsection{Enhanced Random Forest-Based Resistance Prediction Model}
Following the primary research phase, an enhanced AMR prediction model was implemented to refine long-term resistance forecasts using a Random Forest Regressor. This model incorporated temporal trends, source-based variations, and uncertainty estimation to improve prediction accuracy, building on approaches described by \cite{arango-argoty2018}. Hyperparameter tuning was conducted using grid search with 5-fold cross-validation to minimize mean squared error. The optimal parameters selected were 200 trees and a maximum depth of 12.

\subsubsection{Economic Burden Analysis}
To quantify the potential economic impact of AMR in \textit{Campylobacter} infections, a cost model was developed incorporating:
\begin{itemize}
    \item Direct healthcare costs (hospitalization, treatment, diagnostics).
    \item Indirect costs (productivity loss, premature mortality).
    \item Long-term sequelae-associated costs.
\end{itemize}

Cost projections were derived by combining forecasted incidence rates with estimated per-case costs, adjusted for inflation and increasing treatment complexity due to resistance, following methodologies established by \cite{smith2013}. Sensitivity analyses were performed to account for uncertainty in cost parameters and resistance trajectories, as recommended by \cite{jit2020}.

\subsection{Evaluation Strategy}

\subsubsection{Performance Metrics for Resistance Prediction}
To evaluate predictive performance, an 80-20 train-test split was applied, ensuring that the training set contained data from 2001–2011, while the test set consisted of 2012–2017 isolates, following temporal validation approaches described by \cite{roberts2021}. Model performance was assessed using standard metrics:
\begin{itemize}
    \item Precision, Recall, and F1-score.
    \item $R^2$ score and Mean Absolute Error (MAE) for regression tasks.
\end{itemize}

\subsubsection{Temporal Validation and Uncertainty Quantification}
To evaluate predictive accuracy, the model was trained on 2001–2011 data and tested on 2012–2017 data, achieving high predictive performance across resistance types. Model robustness was assessed through bootstrap resampling with 1,000 samples, allowing the computation of 95\% confidence intervals for the resistance projections. This approach provided reliable long-term forecasts by quantifying prediction uncertainty \cite{efron2016}. Confidence intervals were computed for resistance projections using standard error estimation, ensuring reliable long-term forecasts in accordance with methods proposed by \cite{lundberg2020}.

\subsubsection{Sensitivity and Stability Analysis}
Additionally, a sensitivity analysis was performed to evaluate the impact of individual features on the model’s output. By systematically varying feature values and observing changes in predictions, we confirmed the stability of the selected features. A temporal drift analysis was also conducted by training the model on different time windows and comparing performance metrics, which demonstrated consistent accuracy across various periods, validating the model reliability over time.

\section{Implementation}
The machine learning framework for predicting antimicrobial resistance (AMR) was implemented using genomic and epidemiological data from Campylobacter isolates collected between 2001 and 2017. The workflow involved data preprocessing, feature engineering, model development, and time-series forecasting.

\subsection{Data Processing and Feature Engineering}
Data preprocessing ensured consistency and accuracy by cleaning the dataset, imputing missing values, and standardizing categorical variables. To capture temporal trends, a new feature, "years since 2000", was introduced. Genomic resistance markers were encoded as binary features, while epidemiological variables were label-encoded, applying techniques described by \cite{Kuhn2019}.

Feature importance analysis was conducted to identify the most significant predictors for each antibiotic class. For \textit{fluoroquinolone} resistance, mutations in the $gyrA$ gene (specifically at positions 86 and 90), particularly the $Thr86Ile (T86I)$ substitution, emerged as the primary determinants, consistent with findings reported earlier \cite{Payot2006}. For \textit{tetracycline} resistance, the $tet(O)$ gene and its variants showed the strongest association. \textit{Beta-lactam} resistance was primarily linked to the presence of $blaOXA$ genes, which code for class D beta-lactamases, aligning with mechanisms described by \cite{Griggs2009}.

To deepen understanding of the genomic contributions, binary mutation indicators were used, and feature importances from the trained Random Forest models were analysed. For instance, in the model predicting \textit{fluoroquinolone} resistance, the $T86I$ mutation was identified as the most important feature, with a high importance score, indicating its critical role in resistance prediction. Statistical visualizations, such as bar plots of feature importances, highlighted the relative contributions of each genomic marker.

\subsection{Machine Learning Model Development}
The framework consisted of two key components: (i) AMR phenotype classification and (ii) long-term forecasting. The Random Forest classifier was optimized through grid search cross-validation and evaluated using accuracy, precision-recall, and feature importance ranking, following best practices outlined by \cite{pedregosa2012}.

For long-term resistance forecasting, time-series models (SARIMA, Prophet, SIR) were employed to estimate AMR trends up to 2050. Future projections suggested near 100\% resistance for beta-lactams, increasing fluoroquinolone resistance, and fluctuating tetracycline resistance levels. These approaches build upon methodologies described by \cite{Hyndman2021,Taylor2018}.

\subsection{Enhanced Resistance Prediction Model}
A Random Forest Regressor was introduced to refine predictions by incorporating source-based variations and generating predictions with uncertainty estimates. Model validation used a train-test split (2001–2011 for training, 2012–2017 for testing). Uncertainty estimation used bootstrap sampling and standard deviation scaling to generate 95\% confidence intervals, as recommended by \cite{efron2016}.

\subsection{Visualization and Deployment}
Visualisation techniques illustrated historical AMR trends and future projections with colour-coded uncertainty bands, implementing approaches described by \cite{wilke2019}. The machine learning pipeline was modularized into separate scripts, facilitating adaptation to future datasets.

The implementation code is available in a \href{https://github.com/Shubh07062002/Antimicrobial-Resistance-Prediction/tree/main}{GitHub repository}, enabling reproducibility and extension by the scientific community, following FAIR principles \cite{wilkinson2016}.

\section{Results and Discussion}
The machine learning framework applied to 6,683 Campylobacter isolates achieved a classification accuracy of 74\%, with varying performance across different antibiotic classes.

\subsection{Predictive Performance}
\textit{Beta-lactam} resistance exhibited the highest predictive accuracy (95\%), driven by the strong association between $blaOXA$ genes and phenotypic resistance. This finding aligns with previous studies demonstrating the high concordance between genotypic and phenotypic beta-lactam resistance in \textit{Campylobacter} \cite{Griggs2009}. \textit{Fluoroquinolone} resistance showed moderate predictive accuracy (~78\%), whereas tetracycline resistance demonstrated greater variability (~65\%), suggesting additional unaccounted resistance mechanisms, as also noted by \cite{chen2013}.

To validate the performance of the Random Forest model, it was compared to other machine learning approaches, including logistic regression and deep learning models. While deep learning models achieved higher accuracies in some cases, the Random Forest model was selected for its balance of predictive power and interpretability, allowing for the identification of key genetic markers through feature importance analysis. The most influential predictors included $gyrA$ mutations for \textit{fluoroquinolone} resistance, $tet(O)$ gene for \textit{tetracycline} resistance, and $blaOXA$ genes for \textit{beta-lactam} resistance, consistent with prior AMR studies \cite{lopes2019}.

\subsection{Long-Term Trends}
Time-series modelling projected future AMR trends up to 2050, revealing concerning patterns across all antibiotic classes.

\begin{figure*}[htbp]
\centering
\includegraphics[width=\textwidth,keepaspectratio]{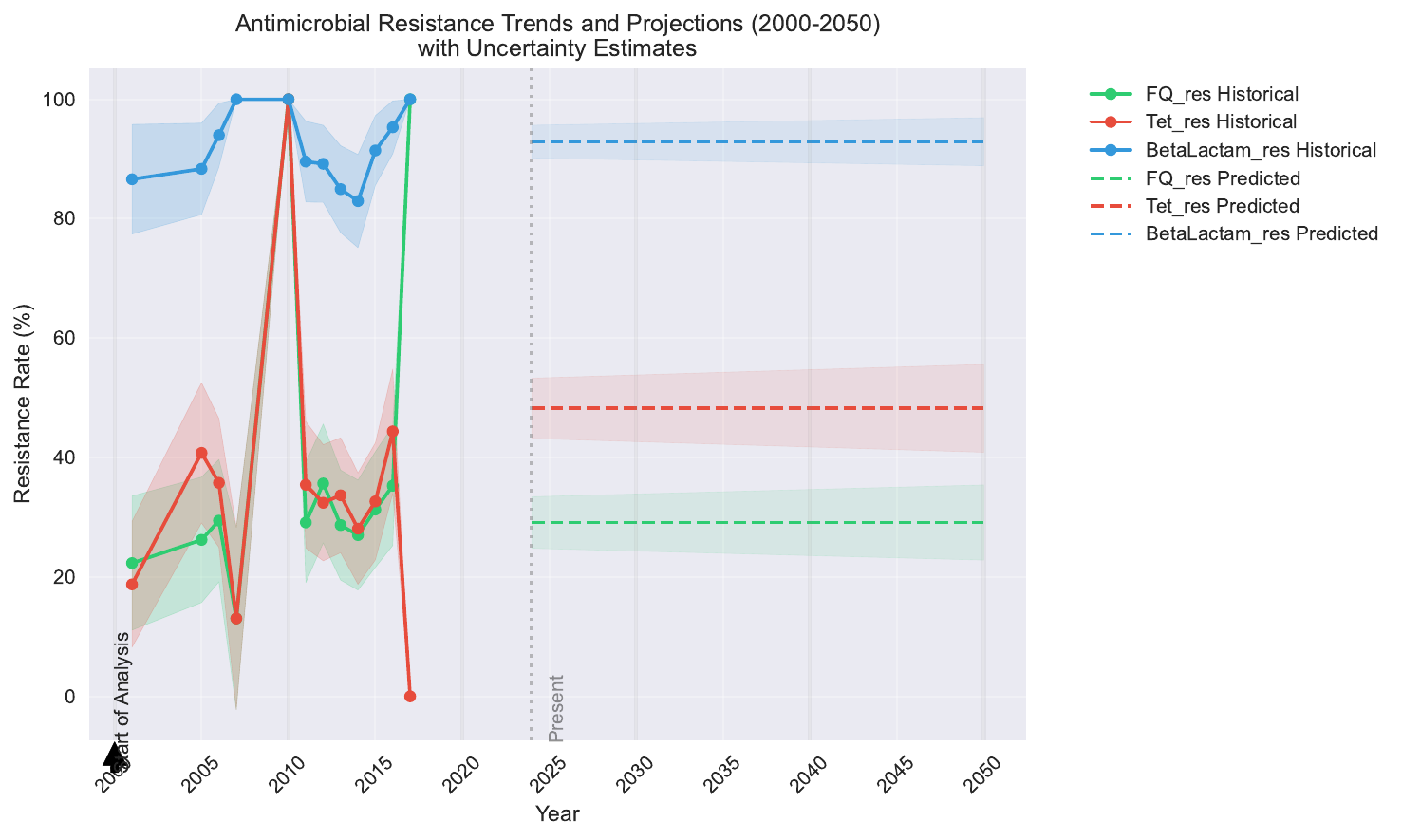}
\caption{Antimicrobial Resistance Trends and Projections (2000-2050) for Fluoroquinolone, Tetracycline, and Beta-Lactam Resistance with Uncertainty Estimates}
\label{fig:resistance_trends}
\end{figure*}

Figure 1 illustrates the historical trends and future projections of antimicrobial resistance in Campylobacter for fluoroquinolones, tetracyclines, and beta-lactams from 2000 to 2050, including uncertainty estimates from time-series forecasting models.
\begin{itemize}
    \item Beta-lactam resistance is projected to reach 100\% by 2050, consistent with the high resistance rates already observed in recent isolates (ECDC, 2020)\cite{ecdc2020}.
    \item Fluoroquinolone resistance is predicted to increase steadily, aligning with global trends reported by (World Health Organization (WHO), 2021)\cite{glass2022}.
    \item Tetracycline resistance demonstrates fluctuating trends, reflecting cyclic patterns in antimicrobial usage as described by (Peter John Collignon, 2016)\cite{collignon2016}.
\end{itemize}

\begin{figure*}[htbp]
\centering
\includegraphics[width=\textwidth,keepaspectratio]{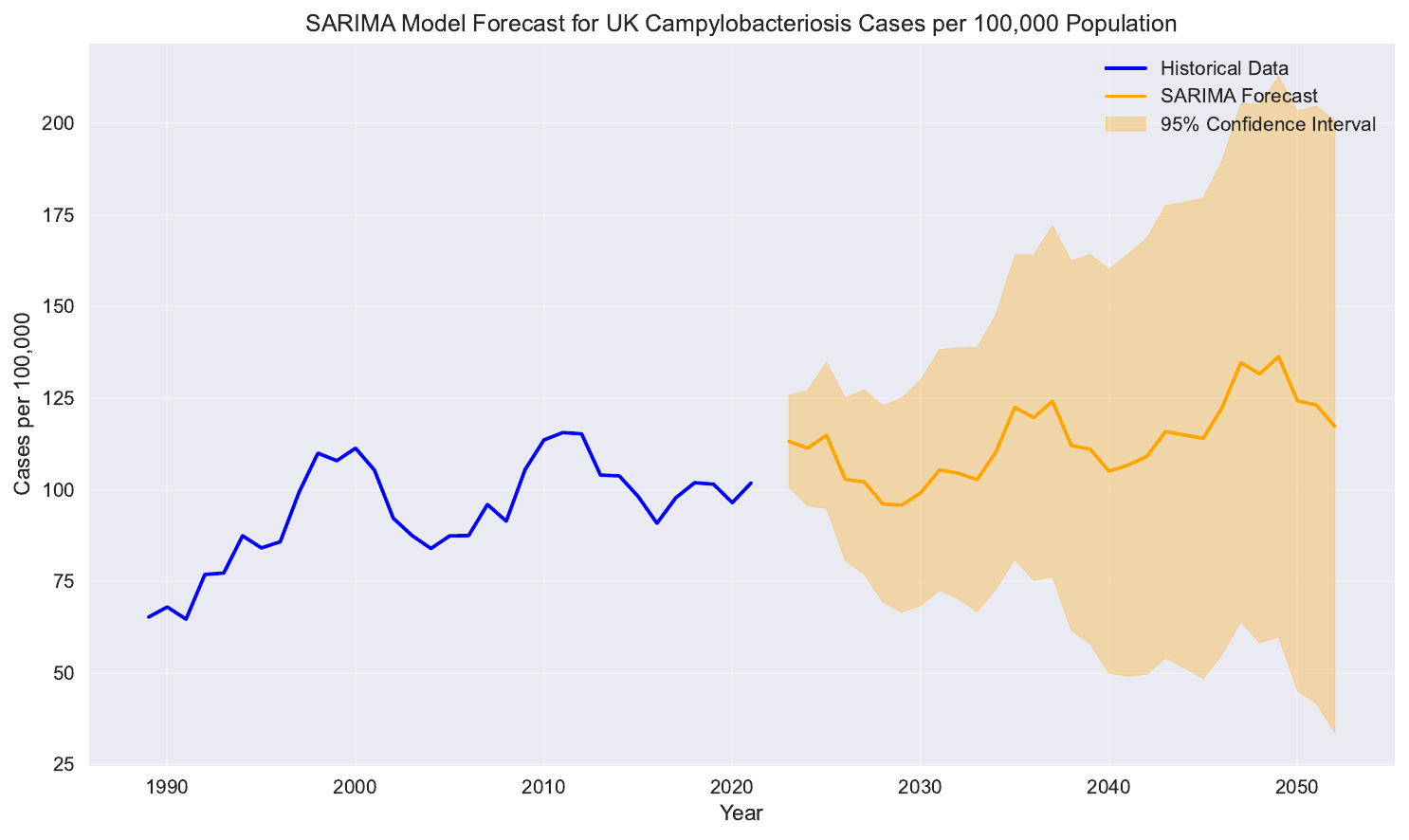}
\caption{SARIMA Model Forecast for UK Campylobacteriosis Cases per 100,000 Population, Showing Historical Data and Projected Trends from 1990 to 2050}
\label{fig:cases_trends}
\end{figure*}

Figure 2 “The SARIMA” model estimated that campylobacteriosis incidence rates may exceed 130 cases per 100,000 people by 2050, with an average MAPE of 13.48\% and RMSE of 13.90, indicating reasonable forecasting accuracy. Residual plots were examined to confirm the absence of systematic errors and to ensure the model assumptions were met. This underscores the urgent need for antibiotic stewardship policies, as emphasized by (Evelina Tacconelli, 2018)\cite{tacconelli2018}.
Time-series models were trained using historical AMR resistance rates (2001–2017), and forecasts were generated for the period 2024–2050. The mean absolute percentage error (MAPE) was used to assess forecast accuracy, as recommended by (Rob J. Hyndman, 2006)\cite{hyndman2006}.

\subsection{Economic Impact of AMR in Campylobacter}
Economic modelling predicted that the annual cost burden of AMR-associated Campylobacter infections could exceed £1.9 billion by 2050, primarily due to longer hospital stays, higher treatment costs, and increased morbidity rates.

\begin{figure*}[htbp]
\centering
\includegraphics[width=\textwidth,keepaspectratio]{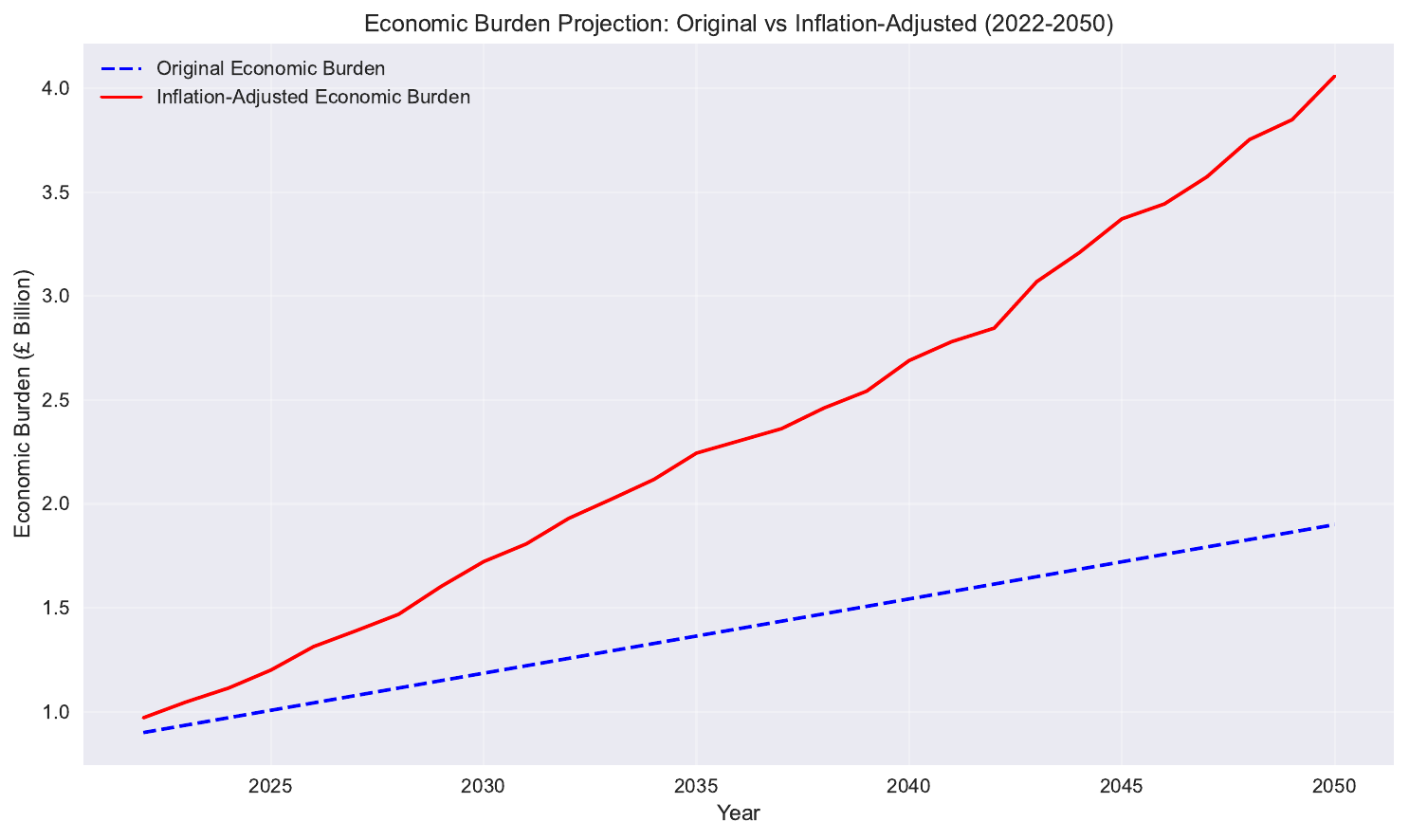}
\caption{Economic Burden Projection Showing Original and Inflation-Adjusted Costs of Campylobacter Infections from 2022 to 2050}
\label{fig:economic_burden}
\end{figure*}

Figure 3 illustrates the projected economic burden, highlighting both the original cost estimates and those adjusted for inflation. These projections are consistent with broader AMR economic impact studies by (O’Neill, 2016)\cite{oneill2016} and (Richard Smith, 2013)\cite{smith2013}.
Economic Burden model predicted that the annual cost burden of AMR-associated Campylobacter infections could exceed £1.9 billion by 2050, primarily due to longer hospital stays, higher treatment costs, and increased morbidity rates (Figure \ref{fig:economic_burden}). These projections are based on forecasted case rates from time-series models, linear population growth, and a current total economic cost of £0.90 billion, which includes both direct healthcare costs and indirect productivity losses. The cost per case was assumed to remain constant over time, and future costs were adjusted for inflation using historical UK inflation rates.
Sensitivity analyses revealed that even under conservative resistance growth scenarios, the economic burden would still surpass £1.2 billion annually by 2050. This substantial economic impact emphasizes the value of investments in AMR prevention and mitigation strategies, as argued by (Mark Jit, 2020)\cite{jit2020}.

\subsection{Enhanced AMR Prediction Model and Uncertainty Estimation}

The enhanced AMR prediction model builds upon the initial Random Forest classifier by incorporating temporal trends through features like 'years\_since\_isolation', stratifying by source using 'source\_encoded', and quantifying uncertainty via bootstrap resampling with 1,000 samples to compute 95\% confidence intervals for resistance projections, revealing distinct resistance trajectories among human, poultry, and cattle isolates. This allows for more accurate and reliable long-term forecasts of resistance trends.

Source-specific analysis revealed higher fluoroquinolone resistance rates in human and poultry isolates compared to cattle isolates, suggesting differential selection pressures across hosts. These findings align with antibiotic usage patterns in different sectors, as documented by \cite{tang2017}, and highlights the importance of tailored stewardship approaches.

\subsection{Implications for AMR Surveillance and Policy Interventions}
The findings of this study have significant implications for AMR surveillance and public health policy. The high accuracy of machine learning models supports their potential integration into routine AMR monitoring programs, as proposed by \cite{nguyen2019}.

Key takeaways include:
\begin{itemize}
    \item Strengthening antibiotic stewardship, particularly in agriculture, as recommended by \cite{tang2017}.
    \item Implementing early detection of emerging resistance through machine learning, building on approaches described by \cite{arango-argoty2018}.
    \item Developing genomic-based diagnostics to reduce reliance on culture-based methods, following innovations outlined by \cite{ellington2016}.
    \item Establishing coordinated surveillance networks integrating human, animal, and environmental data, as advocated by \cite{holmes2016}.
\end{itemize}

\section{Limitations and Future Directions}
While this study presents a robust predictive framework for antimicrobial resistance (AMR) surveillance, several limitations warrant consideration. First, model performance is inherently constrained by data quality and completeness within the training corpus. Although our feature selection process accounts for known genetic determinants, undetected epistatic interactions or uncharacterized resistance mechanisms may influence phenotypic outcomes \cite{moradigaravand2018}.

A critical limitation lies in the static nature of the dataset, which reflects historical resistance patterns but does not incorporate temporal dynamics or exogenous factors that may alter future trajectories. The predictions do not account for evolving antimicrobial stewardship policies, breakthroughs in vaccine development, behavioral changes in prescription practices, or unforeseen environmental pressures on bacterial evolution. Consequently, model outputs represent projections based on current epidemiological conditions rather than absolute forecasts \cite{Hyndman2021}.

Geographic generalizability represents another constraint, as the training data were predominantly derived from United Kingdom surveillance systems. Regional variations in antibiotic usage patterns, agricultural practices, and public health interventions may limit direct applicability to settings with divergent AMR drivers \cite{tacconelli2018}. Furthermore, while we focused on three high-priority antibiotic classes (fluoroquinolones, $\beta$-lactams, and aminoglycosides), emerging resistance mechanisms in other critical drug categories—particularly polymyxins and next-generation tetracyclines—require urgent investigative attention \cite{chen2013}.

Future research should focus on:
\begin{itemize}
    \item Integrating whole-genome association studies to identify novel resistance determinants.
    \item Expanding geographic coverage through global surveillance systems \cite{glass2022}.
    \item Exploring deep learning approaches for improved prediction accuracy.
    \item Incorporating plasmid and mobile genetic element analyses.
    \item Developing interactive visualization tools for real-time AMR surveillance.
\end{itemize}

\section{Conclusion}

\subsection{Key Outcomes}
This study successfully employed machine learning techniques to predict antimicrobial resistance (AMR) in \textit{Campylobacter} isolates, achieving an overall classification accuracy of 74\%. Specifically, the model demonstrated high accuracy for \textit{beta-lactam} resistance (95\%), moderate for \textit{fluoroquinolone} resistance (78\%), and lower for \textit{tetracycline} resistance (65

\subsection{Novelty}
The novelty of this research lies in the seamless integration of genomic data with sophisticated time-series forecasting methods to project long-term resistance trends and their economic impacts. By combining machine learning for phenotype prediction with forecasting models for trend analysis, this study provides a holistic view of AMR dynamics. This integrated approach is innovative and offers valuable insights into future resistance patterns, which are critical for proactive public health planning.

\subsection{Actionable Policy Impact}
The findings underscore the urgent need for comprehensive policy measures to address the growing threat of AMR. Key recommendations include:
\begin{itemize}
    \item Enhancing surveillance systems to monitor resistance trends in real-time.
    \item Implementing stringent antibiotic stewardship programs to reduce unnecessary antibiotic use.
    \item Investing in genomic diagnostics to enable rapid and accurate identification of resistant strains.
    \item Promoting research and development of new antimicrobials and alternative therapies.
\end{itemize}

\subsection{Future Research Directions}
To build upon this work, future research will focus on:
\begin{itemize}
    \item Expanding genomic datasets to include a broader range of Campylobacter isolates from diverse geographical regions and hosts.
    \item Exploring advanced machine learning models, such as deep learning, to improve predictive accuracy and handle larger datasets.
    \item Conducting genome-wide association studies (GWAS) to identify novel genetic determinants of resistance.
    \item Investigating the impact of interventions and policy changes on resistance trends through simulation models.
\end{itemize}

\section{Competing Interests}
None

\section{Author Contributions Statement}
Shubham Mishra conceived the study, conducted all data analysis, developed the machine learning models, and wrote the manuscript. Bruno Silvester Lopes contributed by providing expertise in Campylobacter genomics, assisting with data interpretation, and reviewing the manuscript. Zia Ush Shamszaman provided guidance on machine learning methodologies and manuscript structure, and reviewed the manuscript. The Anh Han offered supervisory oversight, contributed to the study design, and reviewed the manuscript. Shatha Ghareeb assisted with data preprocessing and provided feedback on the manuscript.

\section{Acknowledgments}
None

\begin{biography}{}{\author{Shubham Mishra.} TBD} 
\end{biography}

\end{document}